\documentclass[11pt,twoside]{article}

\usepackage{asp2006}
\usepackage{epsf}
\usepackage{graphicx}
\usepackage{psfig}
\usepackage{lscape}

\begin{document}
\title{The Star Formation History of the Universe}
\author{Andrew M. Hopkins}
\affil{School of Physics, University of Sydney, NSW 2006, Australia}

\begin{abstract}
Strong constraints on the cosmic star formation history (SFH) have recently
been established using ultraviolet and far-infrared measurements, refining
the results of numerous measurements over the past decade. Taken together,
the most recent and robust data indicate a compellingly consistent picture of
the SFH out to redshift $z \approx 6$, with especially tight constraints for
$z < 1$. There have also been a number of dedicated efforts to
measure or constrain the SFH at $z \approx 6$ and beyond. It is also
possible to constrain the normalisation of the SFH using a combination
of electron antineutrino flux limits from Super-Kamiokande measurements
and supernova rate density measurements. This review presents the latest
compilation of SFH measurements, and summarises the corresponding
evolution for stellar and metal mass densities, and supernova rate densities.
The constraints on the normalisation of the cosmic SFH, arising from
the combination of the supernova rate measurements and the measurement limit
on the supernova electron antineutrino flux, are also discussed.
\end{abstract}

\section{Introduction}

In the past few years measurement of the evolution of galaxy luminosity
functions at a broad range of wavelengths has rapidly matured. One consequence
of this has been the refinement in our understanding of how the space density
of the galaxy star formation rate (SFR) evolves \citep{Lil:96,Mad:96}.
In particular the cosmic star formation history (SFH) is now quite tightly
constrained (to within $\approx30-50\%$) up to redshifts of $z\approx1$.
Combined with measurements at higher redshifts the SFH is
well determined (within a factor of about 3 at $z\ga 1$) up to
$z\approx6$ \citep[e.g.,][]{Hop:04}.

Additional results from the Super-Kamiokande (SK) particle detector provide
a strong limit on the electron antineutrino ($\overline{\nu}_e$) flux,
1.2\,cm$^{-2}$\,s$^{-1}$ (for $E_{\nu}>19.3\,$MeV), originating from supernova
type~II events associated with the SFH \citep{Mal:03}. This limit on the
diffuse supernova neutrino background (DSNB) constrains the normalisation of
the SFH \citep{FK:03,Str:05}. An exploration of quantities predicted
from the SFH, the stellar and metal mass density evolution, and supernova (SN)
rate evolution, provides further insight into the allowable normalisation of
the SFH \citep{Str:05}. This series of interconnected physical properties of
galaxies and SNe provides an emerging opportunity for determining
the level of the SFH normalisation, and the SFH measurements particularly
for $z\la 1$ now have the precision to allow this exploration of
their accuracy. Constraining the normalisation of the SFH will support
a range of quantitative analyses of galaxy evolution, including the
mass-dependence of the SFH \citep[e.g.,][]{Pap:06, Jun:05, Hea:04}, and the
reasons underlying the decline in the SFH to low redshifts
\citep[e.g.,][]{Bell:05}.

The analysis of the constraints on the SFH normalisation are detailed
in \citet{Hop:06}, and here the main arguments are summarised. Since optical
SNII can be hidden from observations by dust obscuration, the present SNII
rate density measurements may merely be lower limits, and serve as a lower
bound on the allowable SFH normalisation. In contrast, since neutrinos
are unaffected by dust, the DSNB provides an absolute upper limit on
the true SNII rate.
We assume
$H_0=70\,$km\,s$^{-1}$\,Mpc$^{-1}$, $\Omega_M=0.3$, $\Omega_{\Lambda}=0.7$.

\begin{figure}
\centerline{\rotatebox{-90}{\includegraphics[width=6.5cm]{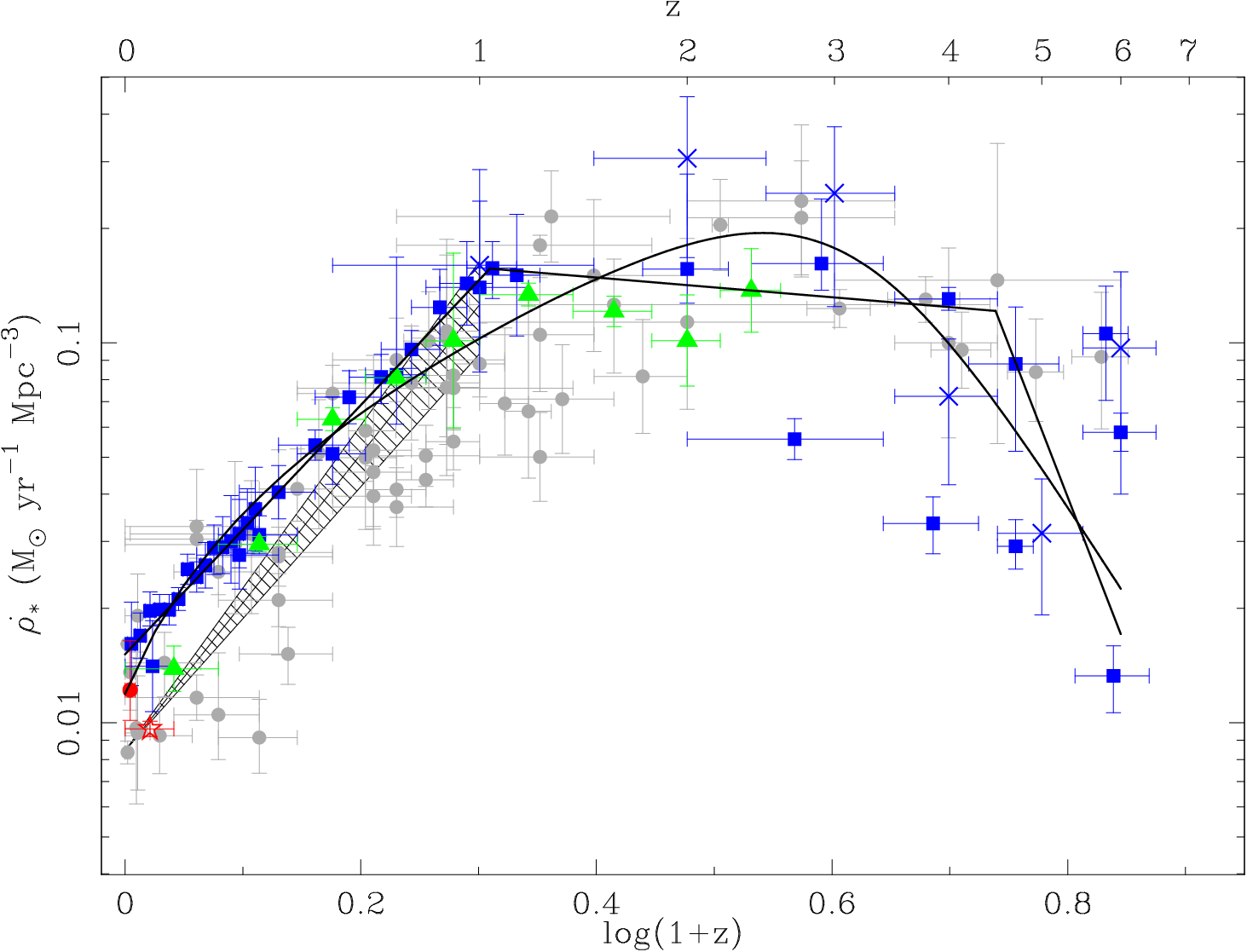}}}
\caption{Evolution of SFR density with redshift (scaled assuming the SalA IMF).
Circles are from the compilation of \citet{Hop:04}. The hatched region is the
$24\,\mu$m SFH from \citet{LeF:05}. Triangles are $24\,\mu$m data from
\citet{Per:05}. The open star at $z=0.05$ is based on 1.4\,GHz data from
\citet{Mau:05}. The filled circle at $z=0.01$
is the H$\alpha$ estimate from \citet{Han:06}. Squares are UV data from
\citet{Bal:05,Wol:03,Arn:05,Bou:03a,Bou:03b,Bou:05,Bun:04,Ouch:04}.
Crosses are the UDF estimates from \citet{Tho:06}.
 \label{fig:sfh}}
\end{figure}

\section{The Data}
The compilation of \citet{Hop:04} was taken as the starting point for this
analysis, and uses their ``common" obscuration correction where necessary.
Additional measurements are indicated in Figure~\ref{fig:sfh}, and are
detailed in \citet{Hop:06}.

\subsection{Dust Obscuration Corrections}

To implement effective obscuration corrections for the UV measurements at
$z\la 1$, we take advantage of the well-established FIR SFR densities up to
$z=1$ from \citet{LeF:05}. The UV data
at $z\le1$ are ``obscuration corrected" by adding the FIR SFR density from
\citet{LeF:05} to each point. As shown by \citet{Bell:03} for individual
systems, this technique results in $\dot{\rho}_*$ estimates consistent with
the obscuration corrected H$\alpha$ measurements. This result
is consistent with the interpretation of \citet{Tak:05} that
about half the SFR density in the local universe is obscured by dust,
increasing to about 80\% by $z\approx1$.
For obscuration corrections to the UV data between $1<z<3$ we rely on
the fact that the FIR measurements of \citet{Per:05} are quite flat in
this domain, as well as being highly consistent with those of \citet{LeF:05}
at $z<1$, and add the constant SFR density corresponding to that
of \citet{LeF:05} at $z=1$. This is also consistent with the recent
measurements of obscuration corrections for UV luminosities at
$z\approx 2$ by \citet{Erb:06}, who find a typical correction factor
of $\approx 4.5$. At higher redshifts we apply a ``common"
obscuration correction to the UV data as detailed in \citet{Hop:04}.

\begin{figure}
\centerline{\rotatebox{-90}{\includegraphics[width=5.0cm]{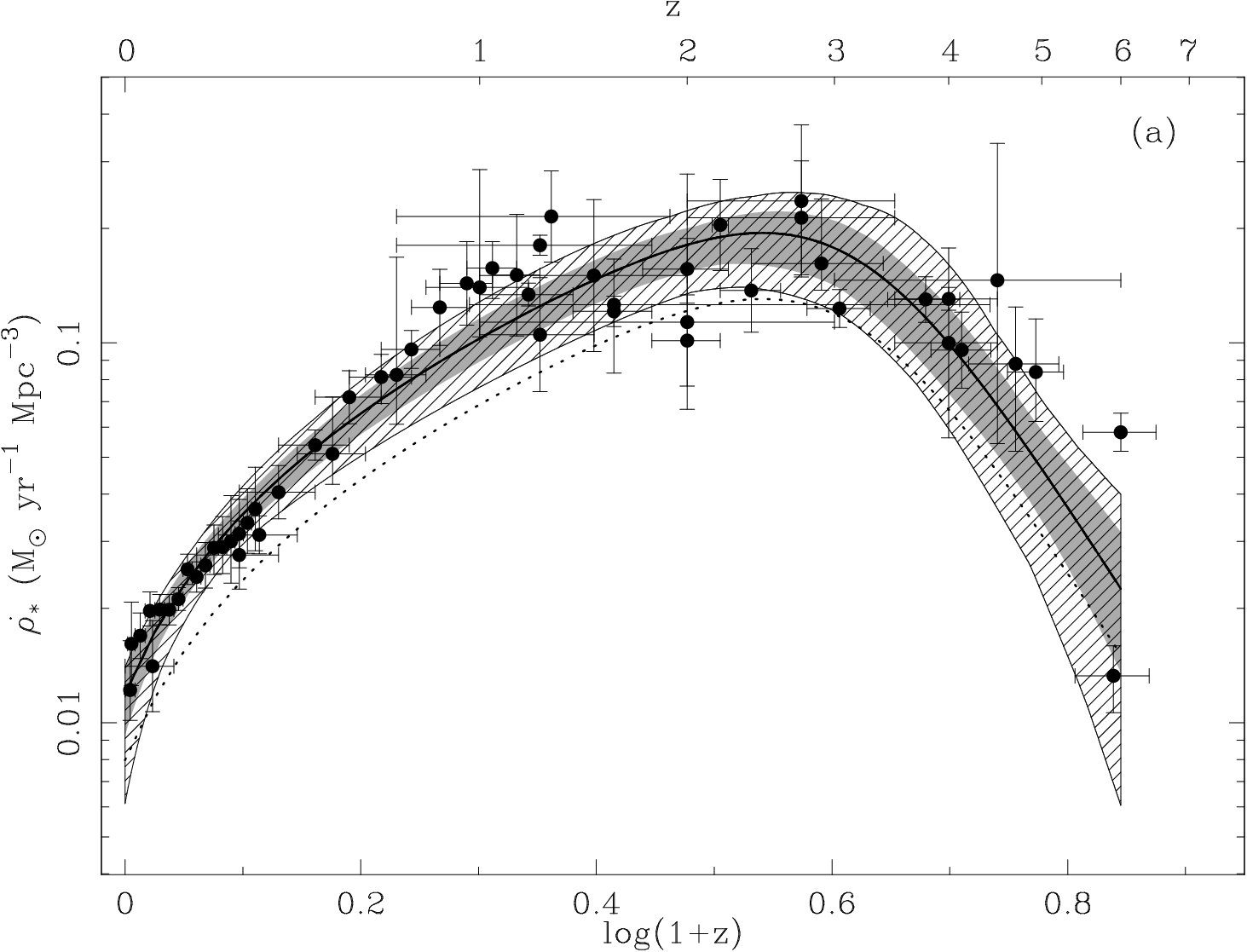}}
\rotatebox{-90}{\includegraphics[width=5.0cm]{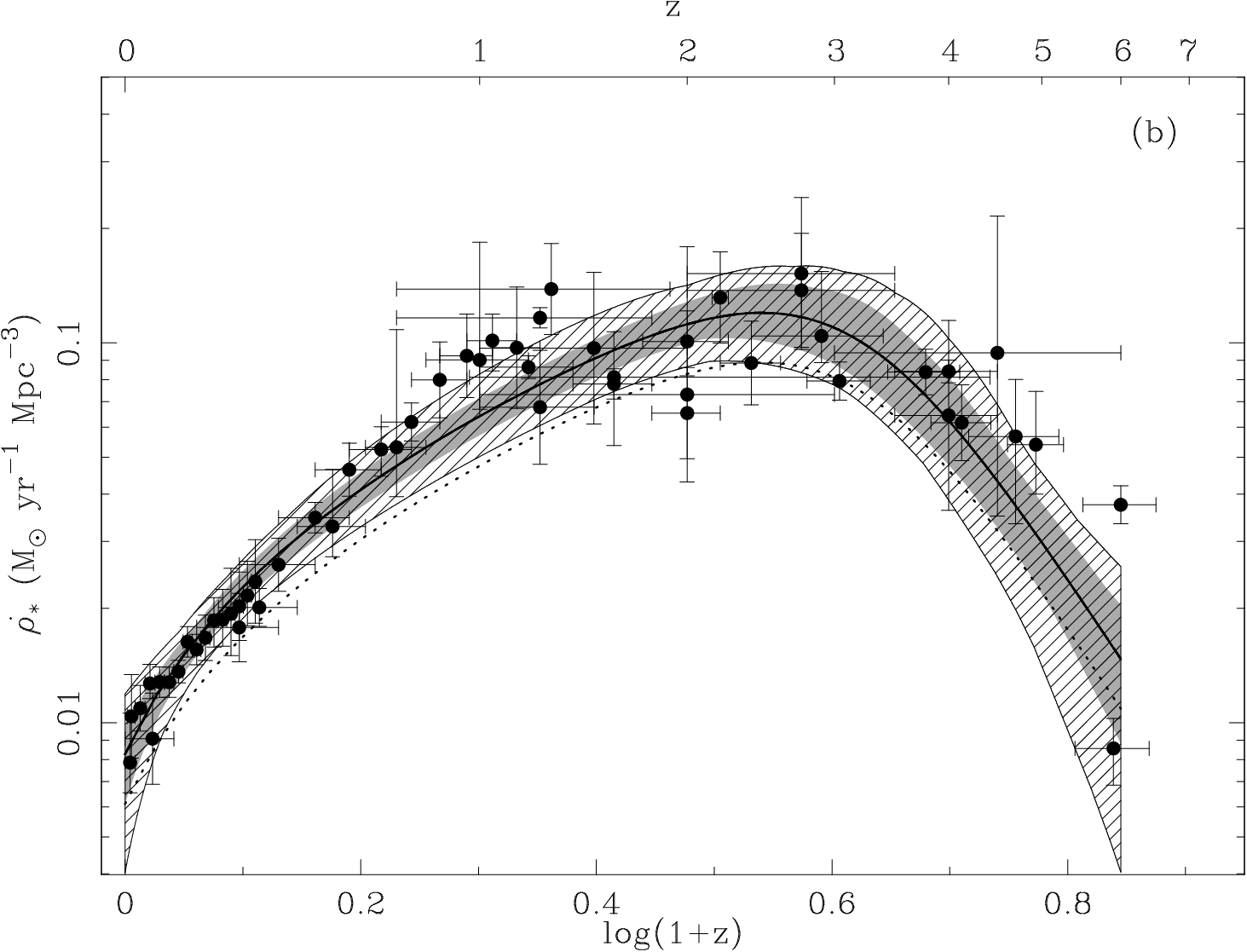}}}
\caption{SFH data used in the parametric fitting.
(a) Assumes SalA IMF. (b) Assumes BG IMF.
The shape of the fits is determined from the SFH data alone, and a scaling
factor is fit to ensure consistency with the SK $\overline{\nu}_e$ limit
Solid lines assume a $\overline{\nu}_e$ temperature of $T=4\,$MeV
or $T=6\,$MeV, and dotted $T=8\,$MeV. The grey shaded and
hatched regions are the $1\,\sigma$ and $3\,\sigma$ confidence regions around
the $T=4\,$MeV fits respectively.
 \label{fig:sfhfit}}
\end{figure}

\subsection{IMF Assumptions}
While uncertainties in SFR calibration act to increase the scatter in
the SFH, and uncertainties in dust obscuration can raise it to greater
or lesser degrees, the choice of IMF is the only assumption that can
systematically {\em decrease\/} the SFH normalisation. A modified
\citet{Sal:55} IMF, with a turnover below $1\,M_{\odot}$, remains a
reasonable model \citep[e.g.,][]{Bal:03}, and other currently favoured
IMFs include those of \citet{Kro:01} and \citet{Bal:03}.
The factor to convert SFH measurements from a traditional \citet{Sal:55} IMF
to the IMF of \citet[hereafter BG IMF]{Bal:03} is $0.50$ ($-0.305\,$dex).
To convert to the modified Salpeter A IMF \citep[hereafter SalA IMF]{Bal:03}
is a factor of 0.77 ($-0.114\,$dex). The \citet{Kro:01} IMF and the modified
Salpeter B IMF \citep[hereafter SalB IMF]{Bal:03} have scale factors
intermediate between these choices. With these two extreme IMF choices
we expect to provide bounds encompassing the result from choosing any
reasonable IMF in our subsequent analysis.

\section{SFH Fitting}
To derive a $\overline{\nu}_e$ flux from the DSNB for comparison
with the limits from SK, we first fit a functional form to the SFH.
We use the parametric form of \citet{Col:01}:
$\dot{\rho}_*=(a+bz)h/(1+(z/c)^d)$, here with $h=0.7$. The individual
$\dot{\rho}_*$ measurements chosen to constrain this fit are also
important since the resulting fit will obviously vary depending on the
data used. The details of the data selected are given in \citet{Hop:06},
and the results shown in Figure~\ref{fig:sfhfit}.
The final parametric fitting is a simple $\chi^2$ fit to the 58 selected
$\dot{\rho}_*$ measurements spanning $0 \le z \la 6$.

\begin{figure}
\centerline{\rotatebox{-90}{\includegraphics[width=7.0cm]{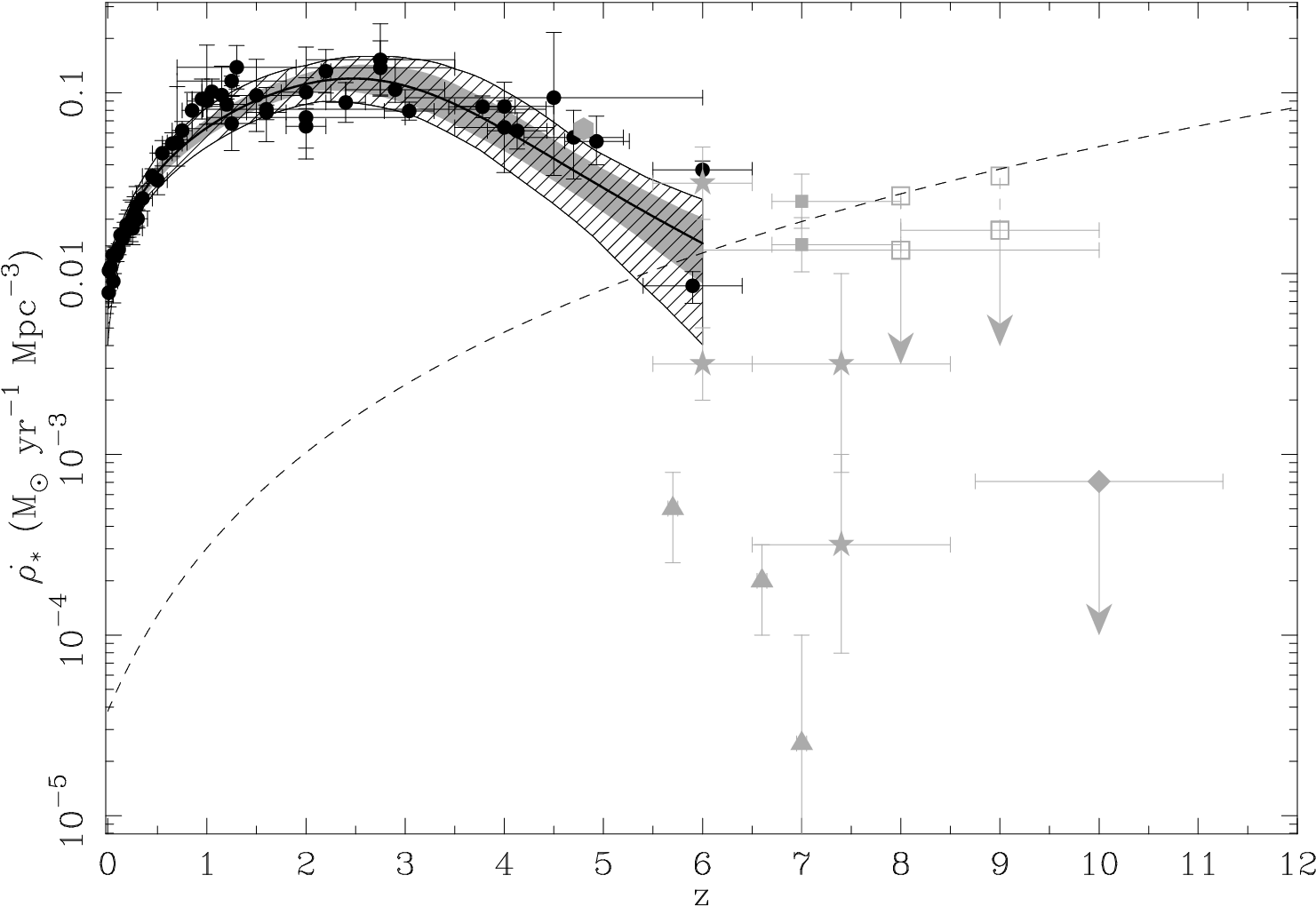}}}
\caption{SFH measurements emphasising recent $z \ga 6$ estimates
(assuming the SalA IMF). The filled circles and grey shaded and hatched
regions are as in the previous Figure (note the different axes scales).
The dashed line corresponds to the level of (unobscured) SFH required for 
reionisation \citep{MHR:99}. Triangles: Ly$\alpha$ emitters \citep{Iye:06};
Stars: LBGs in GOODS \citep{Bou:06}; Filled squares: LBGs in GOODS
\citep{Man:06}; Diamond: LBGs in the HUDF \citep{Bou:05b}; Hexagon: LBGs
(A.\ Verma et al., 2006, this proceedings); Open squares: lensed LBGs
\citep{Ric:06}.
In all cases where the same symbols appear at the same redshift, the upper
values correspond to the maximum expected SFH after integration over the full
luminosity function.
 \label{fig:sfhzgt6}}
\end{figure}

\subsection{UV Data at High Redshift}
As a brief aside, it is worth discussing star formation at $z\ga 6$,
as there are now a large number of estimates in the literature spanning
this range. These primarily use the photometric dropout technique to select
Lyman break galaxies (LBGs), often using relatively small, very
deep fields (including the Great Observatories Origins Deep Survey, GOODS, and
the Hubble Ultra Deep Field, HUDF). Measurements of Ly$\alpha$ emission
typically show SFH measurements significantly lower than those from
LBGs. It is not clear whether the same populations are being
probed here, or the full extent of the selection biases that may be present.

The $z\ga 6$ SFH contributes a negligible fraction to the integrated
SFH that produces the DSNB, and is neglected in the subsequent analysis.
It is, however, of great significance to the epoch of reionisation,
and star formation may in fact play the dominant role here.
Figure~\ref{fig:sfhzgt6} shows the measurements of the $z\ga 6$ SFH,
based on rest-frame UV luminosities not corrected for obscuration effects.
The dashed line in this Figure is the level of SFR density required for
reionisation at a given redshift \citep{MHR:99} assuming $f_{\rm esc}=1$,
appropriate for unobscured UV emission, and a clumping factor $C=30$.

\begin{figure}
\centerline{\rotatebox{-90}{\includegraphics[width=6.0cm]{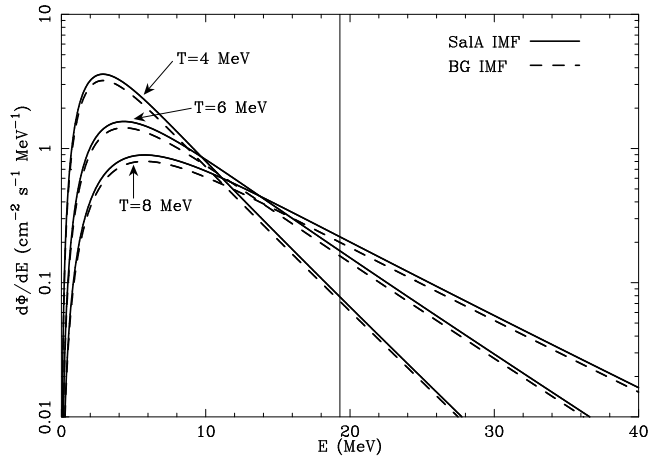}}}
\caption{The $\overline{\nu}_e$ energy spectra predicted for the
various SFH fits and temperature assumptions. The solid and dashed
curves are the SalA IMF and BG IMF respectively. The $T=4\,$MeV and
$T=6\,$MeV curves are consistent with the SK $\overline{\nu}_e$ limit.
The $T=8\,$MeV curves are inconsistent with the $\overline{\nu}_e$ limit,
and indicate the shape of the $\overline{\nu}_e$ spectrum derived by assuming
the parametric form for the SFH corresponding to our best fit ($T=4\,$MeV),
and setting the $\overline{\nu}_e$ temperature to the higher value.
The thin vertical line marks $E=19.3\,$MeV, above which the
$\overline{\nu}_e$ contribute to the SK limit.
 \label{fig:sneas}}
\end{figure}

Although the issue of dust obscuration at high
redshift is still highly uncertain, some data is beginning to be obtained.
In addition to the $E(B-V)$ estimates from \citet{Ouch:04}, intriguing
evidence for significant obscuration ($A_V\approx 1\,$mag) at $z=6.56$ has
recently been established through Spitzer observations of a lensed Lyman
$\alpha$ (Ly$\alpha$) emitting source \citep{Cha:05}. This implies that the
first epoch of star formation in this source must have occurred around
$z\approx 20$. This is also supported by spectroscopic Ly$\alpha$ emission
measurements of LBGs at $z\approx 5$ \citep{And:05},
suggesting that the bright LBGs (at least) lie in dusty, chemically evolved
systems at this redshift. One promising suggestion here is that the
cosmic SFH beyond $z\approx 6$ can be probed through galaxy archeology,
i.e., by determining the star formation histories of the $z\approx 6$
galaxy population (see also R.-R.\ Chary 2006, this proceedings).

\begin{figure}
\centerline{\rotatebox{-90}{\includegraphics[width=4.8cm]{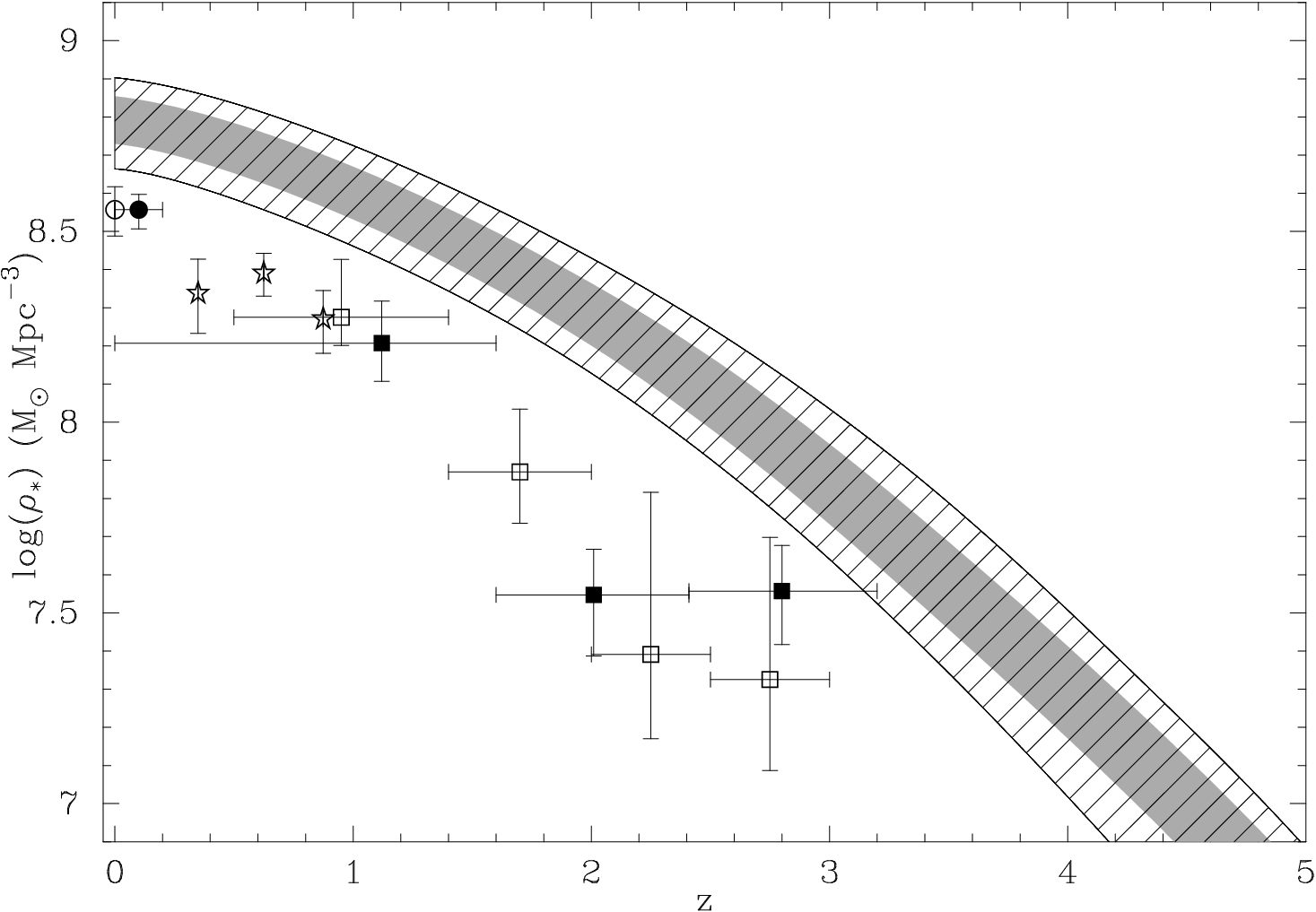}}
\rotatebox{-90}{\includegraphics[width=4.8cm]{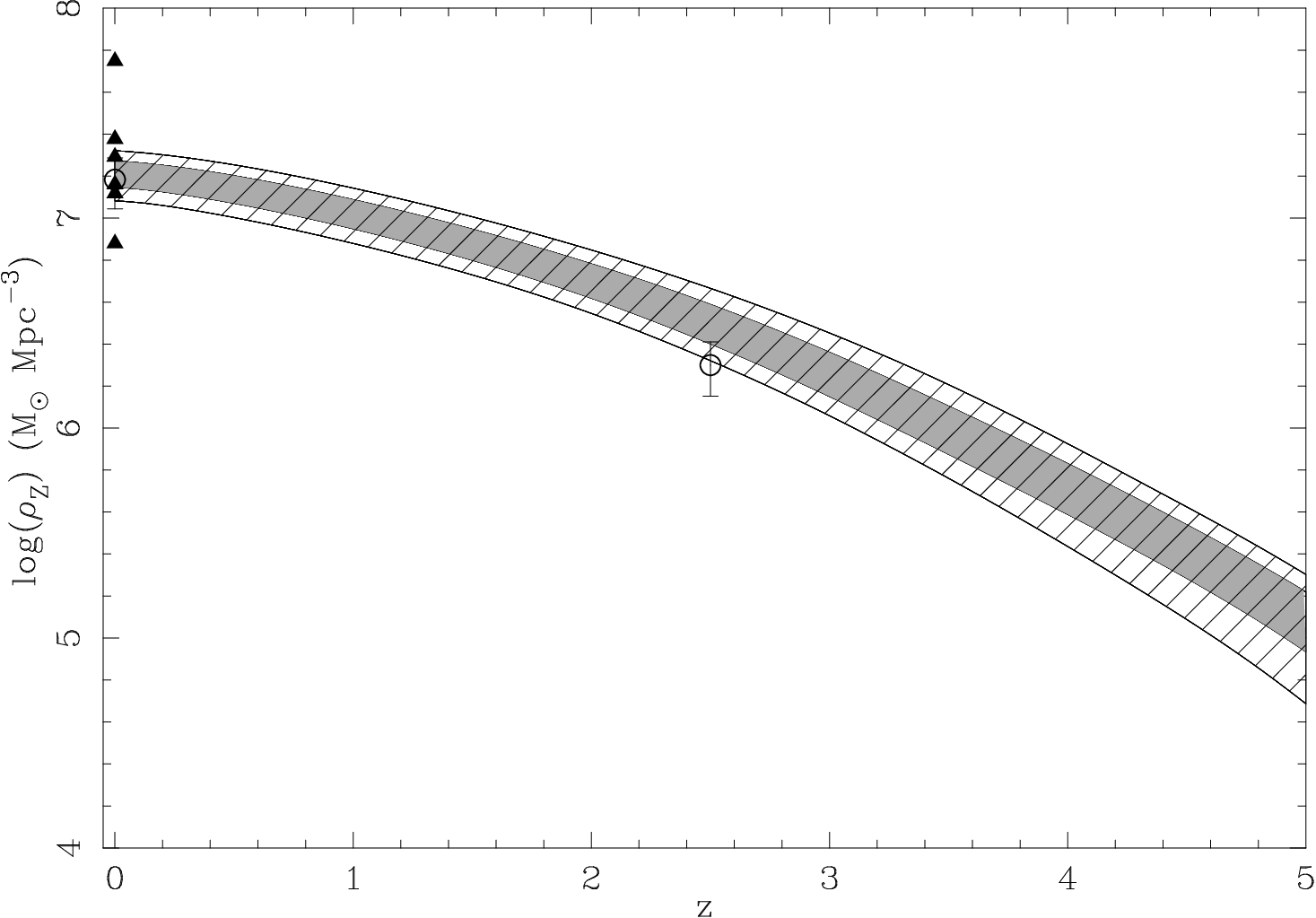}}}
\caption{Left: Evolution of stellar mass density buildup inferred from the SFH.
Right: Evolution of metal mass density buildup inferred from the SFH. Refer
to \citet{Hop:06} for references to the measurements shown in both panels.
The grey shaded and hatched regions come from the $1\,\sigma$ and
$3\,\sigma$ confidence regions around the $T=4\,$MeV fits to the SFH
respectively. The SalA IMF is assumed in all cases.
 \label{fig:stars}}
\end{figure}

\subsection{The Diffuse Supernova Neutrino Background}
The DSNB derived from the cosmic SFH is calculated as follows. The
$\dot{\rho}_*(z)$ is first converted to a type~II supernova rate
history, $\dot{\rho}_{\rm SNII}(z)$. For the IMFs explored here
$\dot{\rho}_{\rm SNII}(z)=(0.0132/M_{\odot})\,\dot{\rho}_*(z)$ for the
BG IMF and $\dot{\rho}_{\rm SNII}(z)=(0.00915/M_{\odot})\,\dot{\rho}_*(z)$
for the SalA IMF.
The predicted differential neutrino flux (per unit energy) is then
calculated by integrating $\dot{\rho}_{\rm SNII}(z)$ multiplied by the
$\overline{\nu}_e$ emission per supernova, appropriately redshifted,
over cosmic time \citep[see][for details]{Hop:06}. Finally, this energy
spectrum is integrated above 19.3\,MeV to establish the $\overline{\nu}_e$
flux for comparison with the SK limit of 1.2\,cm$^{-2}$\,s$^{-1}$
\citep{Mal:03}. We explore the implication of assuming a temperature of
$T\approx4\,$MeV, 6\,MeV, or 8\,MeV.
Given the $\overline{\nu}_e$ flux for each temperature, we simply
scale the best fitting SFH to ensure the SK limit is not violated.
The best fit SFHs independent of the
$\overline{\nu}_e$ limit are identical to the fits constrained by a
$\overline{\nu}_e$ temperature of $T=4\,$MeV. As found and discussed in
\citet{Yuk:05}, our results favour effective temperatures at the lower end of
the predicted range.

In addition to the \citet{Col:01} parameterisation, we also explored a
piecewise linear SFH model in $\log(1+z)-\log(\dot{\rho}_*)$ space,
in order to test the possibility that the \citet{Col:01} parametric model
could be biasing the shape of the resulting SFH fit in some way. In this model
we allow the following six parameters to vary: The $z=0$ intercept, the slopes
of three linear segments and the two redshift values at which the slopes
change.

\section{Results and Discussion}
Figure~\ref{fig:sfh} shows the current SFH data compilation (assuming the
SalA IMF) emphasising the additional data used in this analysis compared to the
compilation of \citet{Hop:04}. The best fitting \citet{Col:01} form
for this IMF is also shown as is the best-fitting piecewise
linear fit. Figure~\ref{fig:sfhfit} shows the data used in the fitting and
the best fits assuming three temperature values for the $\overline{\nu}_e$
population for each IMF assumed. The fitting parameters are tabulated
in \citet{Hop:06}.

\begin{figure}
\centerline{\rotatebox{-90}{\includegraphics[width=6cm]{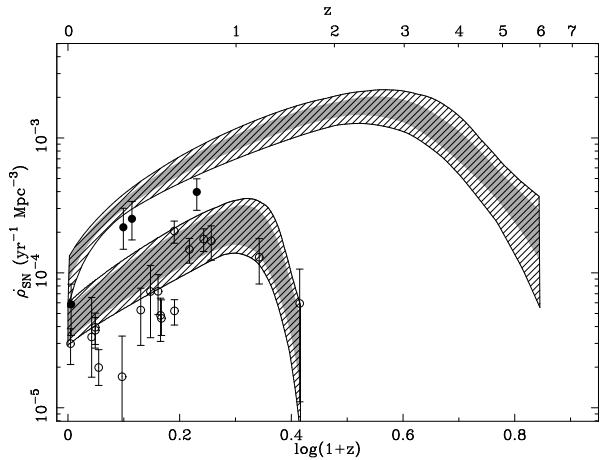}}}
\caption{Evolution of SN rates inferred from the SFH. The upper curves
correspond to the predictions for $\dot{\rho}_{\rm SNII}$, and the lower for
$\dot{\rho}_{\rm SNIa}$, assuming a delay time $t_{Ia}=3\,$Gyr. See
\citet{Hop:06} for references to the measurements shown.
 \label{fig:snrate1}}
\end{figure}

For both assumed IMFs it can clearly be seen that the assumption of
$T=8\,$MeV, when the SFH is required to be consistent with the
$\overline{\nu}_e$ flux limit, is inconsistent with the SFH measurements.
Also for both IMFs, the best fitting SFH assuming $T=6\,$MeV is identical to
that assuming $T=4\,$MeV. This can be understood by considering the SalA IMF,
for example, with the higher SFH normalisation, but which also has a lower
conversion factor between $\dot{\rho}_*$ and $\dot{\rho}_{\rm SNII}$, causing
the predicted $\overline{\nu}_e$ flux to be within the SK limit even with
the assumption of the slightly higher neutrino temperature. For both IMF
assumptions we determine the $1\,\sigma$ (grey-shaded) and
$3\,\sigma$ (hatched) confidence regions around the best fitting SFH
(corresponding to $T=4$ or $6\,$MeV). Subsequent Figures reproduce
these confidence regions in the predictions for stellar and metal mass density
evolution ($\rho_*(z)$ and $\rho_Z(z)$, respectively) and SN rate evolution
($\dot{\rho}_{\rm SN}(z)$).
The shape of the $\overline{\nu}_e$ energy spectra corresponding to the
various fits for the different IMF and temperature assumptions are shown
in Figure~\ref{fig:sneas} \citep[see also][]{BS:06}.

\subsection{Stellar and Metal Mass Densities}

Figure~\ref{fig:stars} shows the evolution of the stellar mass density,
$\rho_*(z)$, \cite[a far more extensive compilation of data appears
in Figure~4 of][]{Far:06}, and the metal mass density, $\rho_Z(z)$,
as inferred from the SFH \citep{PF:95,Mad:96}. The predictions from
the best fitting SFH are also shown.
For the evolution of $\rho_*(z)$ we need to know the
fraction of the stellar mass recycled into the interstellar medium as stellar
winds or SN ejecta, $R$, corresponding to each IMF
\citep{Col:01,Mad:98}. We find
$R=0.40$ for the SalA IMF, and $R=0.56$ for the BG IMF. The stellar mass
inferred is then a fraction $(1-R)$ of the time integral of the SFH
\citep{Col:01}. Converting the observed stellar mass density measurements
(where a Salpeter IMF is assumed) to our assumed IMFs is achieved
by scaling by the product of the SFR conversion factor and the ratio of the
$1-R$ factor for the chosen IMF to that of the Salpeter IMF (where $1-R=0.72$).
The $\rho_*(z)$ measurements in Figure~\ref{fig:stars} clearly lie
systematically below the predictions from the SFH. Simulations suggest
that the measurements might be underestimating the total $\rho_*(z)$
\citep[e.g.,][]{Nag:04,Som:01}. Discussion by \citet{Dic:03}
imply that it is perhaps not unreasonable to expect about a factor of two
larger stellar mass densities as a result of reasonable obscuration levels.
Other issues that have also been raised include incomplete galaxy population
sampling and cosmic variance that may affect surveys probing small fields
of view \citep[see discussion in][]{Nag:04}. At low redshift, the discrepancy
between the measurements of $\rho_*(z)$ and the SFH prediction is more of a
concern, and is discussed further by \citet[and references therein]{Hop:06}.

To determine $\rho_Z(z)$ from the SFH, we assume that
$\dot{\rho}_*=63.7\,\dot{\rho}_Z$ \citep[e.g.,][]{Con:03}. At $z=0$ the
compilation of data from \citet{CM:04} is shown, and these authors favour a
value of $1.31\times10^7\,M_{\odot}$Mpc$^{-3}$. Values at $z=0$ and $z=2.5$
from \citet{Dun:03} are also shown,
suggesting that the evolution in $\rho_Z$ from the SFH may be
consistent with that estimated from the dusty submillimeter galaxy (SMG)
population, although recent results from \citet{BLP:05} indicate that the
SMGs may contribute much less to the metal mass density at high redshift.
Investigating the predictions of $\rho_{Z}(z)$ from the SFH are complicated
by the limited number of estimates for this quantity at $z>0$. This is
observationally a difficult measurement to make, particularly as much of
the metals may exist in an ionised intergalactic medium component
\citep[see, e.g.,][]{Dun:03,BLP:05}.

\subsection{Supernovae Type~Ia and Type~II}

Figure~\ref{fig:snrate1} shows the evolution in the SN rate for both types Ia
and II SNe. The SNII rate density, $\dot{\rho}_{\rm SNII}$, can be calculated
directly from the SFH. The SNIa rate density, $\dot{\rho}_{\rm SNIa}$,
involves more assumptions about the properties of SNIa events than in the
case of SNII. See \citet{Hop:06} for details of these calculations.
The $\dot{\rho}_{\rm SNII}$ measurements provide a strong lower bound on the
normalisation of the SFH. Particularly given that uncertainty regarding
obscuration corrections is more likely to raise than lower the
$\dot{\rho}_{\rm SNII}$ measurements, the SFH normalisation cannot
realistically be much lower than that obtained from assuming the BG IMF
(Figure~\ref{fig:sfhfit}b). Moreover, the $\dot{\rho}_{\rm SNII}$ measurements
are unlikely to be affected by sufficient obscuration to support an SFH
normalisation much higher than that obtained with the SalA IMF \citep{Hop:06}.

The prediction for $\dot{\rho}_{\rm SNIa}$ from the SFH is also particularly
intriguing. The assumption of the fixed $t_{Ia}=3\,$Gyr has the effect of
matching the $z\ga 3$ turnover in the fitted SFH with the apparent
decline in $\dot{\rho}_{\rm SNIa}$ seen in the highest redshift measurement
from the GOODS sample of \citet{Dah:04}. It is possible, indeed probable,
that this is simply a coincidence as it is a single $\dot{\rho}_{\rm SNIa}$
measurement, with large uncertainties, that is suggestive of the decline, and
the turnover in the SFH is driven almost entirely by the $z\approx 6$
measurement of \citet{Bun:04}. It is thus still highly possible that the
decline in both the SFH and $\dot{\rho}_{\rm SNIa}$ lie at somewhat higher
redshift \citep[see][]{Hop:06}.

\section{Summary}

The SFH compilation of \citet{Hop:04} has been updated, emphasising the
strong constraints from recent UV and FIR measurements, and refining
the results of numerous measurements over the past decade.
The results of parametric fits to the SFH, constrained by the
SK $\overline{\nu}_e$ limit, suggest that the preferred IMF
should produce normalisations within the range of those from the
modified Salpeter A IMF \citep{Bal:03} and the IMF of \citet{Bal:03}. They
also suggest that lower temperatures ($T=4-6\,$MeV) are preferred for the
$\overline{\nu}_e$ population.

Based on the fits to the SFH we predict the evolution of $\rho_*$,
$\rho_{Z}$ and $\dot{\rho}_{\rm SN}$, and compare with observations.
The $\dot{\rho}_{\rm SN II}$ measurements provide a key constraint on
the SFH normalisation, and correspondingly on the favoured IMF. In
particular, these data bound the SFH from {\it below}, while the DSNB bounds
the SFH from {\it above}. Together, these provide a novel technique
for testing or verifying measurements of a universal IMF. More measurements of
$\dot{\rho}_{\rm SN}$ for both type~II and Ia SNe over a broader redshift
range would help to more strongly constrain both the
preferred universal IMF and the properties of SNe. Observing the high
redshift turnover in the SNIa rate would also have strong implications for
the location of the expected high redshift turnover in the SFH.

With the best fitting SFH models explored here, the predictions for
the DSNB appear to lie excitingly close to the measured $\overline{\nu}_e$
flux limit. Direct observation of the DSNB will clearly allow much greater
insight into the physics and astrophysics of star formation and supernovae.
Already the DSNB constraint indicates a preferred IMF range and normalisation
for the SFH. It also illustrates that stronger constraints on the SFH have
implications for understanding the details of both SNII and SNIa production,
and the physical basis of neutrino generation by SNII is intimately associated
with all these predictions. Being able to detect the DSNB and its energy
spectrum will allow a more sophisticated analysis of the detailed connections
between all these aspects of star formation and the cosmic SFH.
Methods for increasing the sensitivity of particle detectors to DSNB
antineutrinos and neutrinos have been detailed elsewhere, and
\citet{Hop:06} summarise some of these proposals. 

\acknowledgements
I thank John Beacom for getting me interested in the constraints enabled
through the SK measurements, and for all his help in understanding the
details.
I also acknowledge support provided by the Aus\-tral\-ian Research Council
in the form of a QEII Fellowship (DP0557850).

\end{document}